\documentclass[journal]{IEEEtran}
\usepackage{amsthm,amssymb,graphicx,multirow,amsmath,color,amsfonts}
\usepackage[update,prepend]{epstopdf}
\usepackage[noadjust]{cite}
\usepackage[latin1]{inputenc}
\usepackage{tikz}
\usetikzlibrary{arrows,calc}	
\usepackage{bbm} 
\usepackage{pdfpages}
\usepackage{tabulary}
\usepackage{multirow}
\usepackage{comment}

\def\nbx{{\mathbf{x}}}
\def\nby{{\mathbf{y}}}

\def\nb0{{\mathbf{0}}}
\def\nb1{{\mathbf{1}}}

\def\ncalJ{{\mathcal{J}}}

\def\ncalL{{\mathcal{L}}}

\def\ncalN{{\mathcal{N}}}

\def\ncalR{{\mathcal{R}}}

\def\ncalX{{\mathcal{X}}}

\def\nbbE{{\mathbb{E}}}

\def\nbbP{{\mathbb{P}}}

\def\nbbR{{\mathbb{R}}}

\newtheorem{lemma}{Lemma}

\newtheorem{definition}{Definition}

\newtheorem{theorem}{Theorem}

\newtheorem{remark}{Remark}

\allowdisplaybreaks 
\begin{document}
\graphicspath{{./Figures/}}
\title{
Fundamentals of Drone Cellular Network Analysis under Random Waypoint Mobility Model 
}
\author{
Morteza Banagar and Harpreet S. Dhillon
\thanks{The authors are with Wireless@VT, Department of ECE, Virginia Tech, Blacksburg, VA (email: \{mbanagar, hdhillon\}@vt.edu). The support of the US NSF (Grant CNS-1617896) is gratefully acknowledged.}
}

\maketitle

\begin{abstract}
In this paper, we present the first stochastic geometry-based performance analysis of a drone cellular network in which drone base stations (DBSs) are initially distributed based on a Poisson point process (PPP) and move according to a random waypoint (RWP) mobility model. The serving DBS for a typical user equipment (UE) on the ground is selected based on the nearest neighbor association policy. We further assume two service models for the serving DBS: (i) {\em UE independent model} (UIM), and (ii) {\em UE dependent model} (UDM). All the other DBSs are considered as interfering DBSs for the typical UE. We introduce a simplified RWP (SRWP) mobility model to describe the movement of interfering DBSs and characterize its key distributional properties that are required for our analysis. Building on these results, we analyze the interference field as seen by the typical UE for both the UIM and the UDM using displacement theorem, which forms the basis for characterizing the average rate at the typical UE as a function of time. To the best of our knowledge, this is the first work that analyzes the performance of a mobile drone network in which the drones follow an RWP mobility model on an infinite plane.
\end{abstract}


\section{Introduction} \label{sec:Intro}
Mobility of wireless nodes is known to have a fundamental impact on the performance of wireless networks \cite{J_Grossglauser_Mobility_2002}. Not surprisingly, past decades have seen a significant amount of research on characterizing the performance of a variety of wireless networks under several standard mobility models, such as random walk (RW) or RWP \cite{J_Bandyopadhyay_Stochastic_2007, J_Bettstetter_Node_2003}. However, a common feature of all these works is the assumption that the base stations (BSs) or access points are static while the UEs are mobile. Although this was of course reasonable for conventional terrestrial networks, it is no longer accurate for drone-assisted communication networks, where the drones may act as BSs \cite{J_Zeng_Wireless_2016, J_Mozaffari_Tutorial_2018}. Inspired by this important use case, there has been some recent interest in incorporating drone mobility in the system-level performance analyses of drone-assisted cellular networks. Since this line of work is still in its nascent stages, some fundamental problems, such as the analysis of drone networks under RWP mobility model, are still open. Inspired by this, the main objective of this paper is to present a complete analysis of drone network performance under RWP mobility that involves the novel characterization of the time-varying interference field and consequently the average link rate as a function of time.

\emph{Related Works.}
Given its ability to capture irregularity in the drone locations, there has been an increasing interest in using stochastic geometry for the system-level analysis of drone networks. The authors of \cite{J_Chetlur_Downlink_2017} considered a finite network of static DBSs distributed as a binomial point process (BPP) and analyzed the coverage probability of the network. In \cite{J_Wang_Modeling_2018}, a superposition of macro and aerial BSs is considered in which probabilistic line-of-sight (LoS) and non-line-of-sight (NLoS) propagation models were adopted for the channel. The authors of \cite{J_Enayati_mobile_2018} added mobility to the BPP-modeled DBS network of \cite{J_Chetlur_Downlink_2017} and designed stochastic trajectory processes for the mobility of DBSs in order to gain the same coverage profile as the static case, while improving the average fade duration. Analysis of the link capacity between drones was performed in \cite{J_Yuan_Capacity_2018}, where the authors characterized the distance distribution between drones with random 3D trajectories. A comprehensive survey on mobility in cellular networks, including drone networks, has been recently done in \cite{J_Tabassum_Fundamentals_2019}, where the authors provide an in-depth tutorial on mobility-aware performance analysis of these networks. In \cite{J_Sharma_Coverage_2019} and \cite{J_Sharma_Random_2019}, the analysis of coverage probability is performed for a finite 3D network of mobile DBSs, in which the serving DBS is assumed to hover at a fixed location above the UE \cite{J_Sharma_Coverage_2019} or move \cite{J_Sharma_Random_2019}, while the interfering DBSs follow RWP and RW mobility models for vertical and horizontal displacements, respectively. Continuing on the same general direction, this paper presents the first comprehensive analysis of drone networks under SRWP mobility model. Our main contributions are summarized next.


\emph{Contributions and Outcomes.}
We consider a mobile network of DBSs operating at a specific height above the ground with the initial locations of the DBSs being modeled as a homogeneous PPP. This drone network is assumed to serve UEs on the ground (modeled as an independent PPP). We assume the nearest neighbor association policy to determine the serving DBS for the typical UE, while all the other DBSs act as interfering DBSs. For the mobility of the interfering DBSs, we introduce the SRWP mobility model as a special case of the RWP mobility model and accurately characterize its relevant distributional properties. We then propose two service models for the serving DBS, i.e. (i) UIM, where the serving DBS moves based on the SRWP mobility model, and (ii) UDM, where the serving DBS approaches the typical UE at a constant height and keeps on hovering above the location of the typical UE until its transmission to the typical UE is complete. For this mobility model, we characterize the interference field as seen by the UE for both the service models and show that the network of interfering DBSs for the UIM is an inhomogeneous PPP which does not vary over time, while it is a time-varying inhomogeneous PPP for the UDM. We finally compute the received rate for the UE for both service models. The non-linear mobility model for the drones considered in this paper significantly generalizes the ``straight-line" mobility models considered by the standardization bodies, such as the third generation partnership project (3GPP).


\section{System Model} \label{sec:SysMod}
We consider a network of mobile DBSs operating at a height $h$ from the ground. We model the temporal evolution of the projections of the DBS locations on the ground as the sequence of point processes $\Phi_{\rm D}(t) \subset \nbbR^2$, where $t\in \nbbR^+$ denotes the time index. We assume that the projections of the initial locations of the DBSs on the ground are distributed as a homogeneous PPP with density $\lambda_0$, i.e., $\Phi_{\rm D}(0) \sim {\rm PPP}(\lambda_0)$. This drone network is assumed to serve UEs on the ground that are assumed to be distributed as an independent PPP $\Phi_{\rm U} \subset \nbbR^2$. We assume that the origin ${\bf o} = (0,0,0)$ of the 3D coordinate system is located on the ground and the $xy$-plane is aligned with the ground. Throughout this paper, we refer to the $z=h$ plane as the DBS plane and the projection of the origin onto this plane as ${\bf o'} = (0, 0, h)$. Our focus will be on the downlink analysis for the typical UE, which can be placed at the origin without loss of generality. We denote the distance of a DBS at time $t$ located at $\nbx(t)\in\Phi_{\rm D}(t)$ from $\bf{o'}$ and $\bf{o}$ by $u_\nbx(t) = \|\nbx(t) - \bf{o'}\|$ and $r_\nbx(t) = \sqrt{u_\nbx(t)^2+h^2}$, respectively. Furthermore, the location of the nearest DBS to $\bf{o'}$ and its distances to $\bf{o'}$ and $\bf{o}$ at time $t$ are denoted by $\nbx_0(t)$, $u_0(t)$, and $r_0(t) = \sqrt{u_0(t)^2+h^2}$, respectively. For notational simplicity, we drop the time index $t$ for the distances defined at $t=0$, i.e., $u_0 \triangleq u_0(0)$, $r_0 \triangleq r_0(0)$, $u_\nbx \triangleq u_\nbx(0)$, and $r_\nbx \triangleq r_\nbx(0)$.


In this paper, we assume a nearest neighbor association policy in which the typical UE connects to its closest DBS at every time $t$. We term this DBS as the {\em serving DBS} for the typical UE while all the other DBSs act as interfering DBSs for this UE. We assume that interfering DBSs follow an SRWP mobility model defined as follows.
\begin{definition} \label{SRWP}
\emph {(SRWP).} In the beginning, each DBS hovers for a fixed time $w$ (termed as ``hover time") at its initial location $\nbx(0)$. It then selects a uniformly random direction $\theta \sim U[0, 2\pi)$, independently of the other DBSs, and moves a fixed distance $s$ (termed as ``flight distance") in this direction with a constant velocity $v$ to arrive at $\nbx(w + \frac{s}{v})$. At this location, it hovers for $w$ before moving a distance of $s$ in another random direction and repeats this procedure.
\end{definition}
Just to put the generality of this model in context, note that the state-of-the-art mobility model used by the standardization bodies, such as 3GPP, assumes that each drone moves on a straight line without stopping and changing its direction \cite{3gpp_36777}. Therefore, one can easily argue that such ``straight-line" mobility models are special cases of the model used in this paper. We further consider two service models for the serving DBS defined as follows.
\begin{enumerate}
\item UIM: The serving DBS follows an SRWP mobility model, independent of the typical UE location.
\item UDM: The serving DBS moves towards $\bf{o'}$ in the DBS plane and keeps hovering at this location until its transmission to the typical UE is completed.
\end{enumerate}
While handover may occur in the UIM, that is not a possibility in the UDM if all DBSs have the same velocity. Furthermore, UDM is the best-case model from the perspective of minimizing the serving distance between the typical UE and its serving DBS. It is fair to say that the reality will lie somewhere in between the UIM and the UDM based on the mission requirements of the drones.



We define the received signal-to-interference ratio (${\rm SIR}$) at time $t$ as
\begin{equation} \label{Eq:SIR}
{\rm SIR}(t) = \frac{P h_0(t) r_0(t)^{-\alpha}}{I(t)},
\end{equation}
where $P$ is the DBS transmit power, which is assumed to be equal for all DBSs at all times, $h_0(t)$ is the fading power gain between the serving DBS and the typical UE, $\alpha$ is the path loss exponent, and $I(t)$ is the interference power defined as $I(t)=\sum_{\nbx(t) \in \Phi_{\rm D}'(t)}P h_\nbx(t) r_\nbx(t)^{-\alpha}$, where $\Phi_{\rm D}'(t) \equiv \Phi_{\rm D}(t)\backslash \nbx_0(t)$ represents the point process of the interfering DBSs and $h_\nbx(t)$ is the fading power gain between the interfering DBSs and the typical UE. Rayleigh fading is assumed with a mean power of $1$, which is justified when there is rich local scattering around the typical UE. This gives $h_0(t) \sim {\rm exp}(1)$ and $h_\nbx(t) \sim {\rm exp}(1)$. Note that in the UDM, we have $u_0(t) = [u_0 - vt]^+$, where $[a]^+=a$ if $a\geq 0$ and $[a]^+=0$ otherwise.

The network performance under both service models will be characterized in terms of the average rate achieved by the typical UE at time $t$. This metric is defined as $R(t) = \nbbE[\log\left(1 + {\rm SIR}(t)\right)]$, where the expectation is taken over the PPP $\Phi_{\rm D}(t)$ and the trajectories. This is essentially the average rate experienced by the typical UE at time $t$ across different network and trajectory realizations.


\section{SRWP Interference Field Characterization} \label{sec:Density}
We start our analysis by characterizing the density of the network of interfering DBSs for both service models. Next lemma is the direct consequence of the {\em displacement theorem} because of which it is stated without a proof \cite{B_Haenggi_Stochastic_2012}.
\begin{lemma} \label{lem:NoExclusion}
Let $\Phi$ be a PPP with density $\lambda_0$. If all the points of $\Phi$ are displaced independently of each other and their displacements are identically distributed, then the displaced points form another PPP with the same density $\lambda_0$.
\end{lemma}
In the UIM, since all the DBSs (including the serving DBS) are displaced based on the SRWP mobility model, we can infer from Lemma \ref{lem:NoExclusion} that the network of DBSs at any time $t$ will remain a PPP with density $\lambda_0$. Consequently, the {\em interfering} DBSs will follow an inhomogeneous PPP with density
\begin{equation} \label{LambdaServiceModel1}
\lambda(t; u_\nbx, u_0)=\left\{\begin{matrix}
\lambda_0 & u_\nbx > u_0(t)\\ 
0 & u_\nbx \leq u_0(t)
\end{matrix}.\right.
\end{equation}
Note that although $u_0(t)$ in \eqref{LambdaServiceModel1} varies over time, its distribution does not change. The characterization of $\Phi_{\rm D}'(t)$ for the UDM is not so straightforward and will be the main focus of the rest of this section. It is clear from our construction that $\Phi_{\rm D}'(0)$ is an inhomogeneous PPP with density given by \eqref{LambdaServiceModel1}, which introduces an \emph {exclusion zone}, $\ncalX = b({\bf o'}, u_0)$, for the interfering DBSs, where $b({\bf o}, r)$ is a disc of radius $r$ centered at ${\bf o}$. One can argue directly using displacement theorem that $\Phi_{\rm D}'(t)$ remains an inhomogeneous PPP for the UDM as well. However, unlike the UIM, the characterization of the density of $\Phi_{\rm D}'(t)$ requires some effort. This result is the main focus of the next lemma. 

\begin{figure}
\centering
\includegraphics[width=0.8\columnwidth]{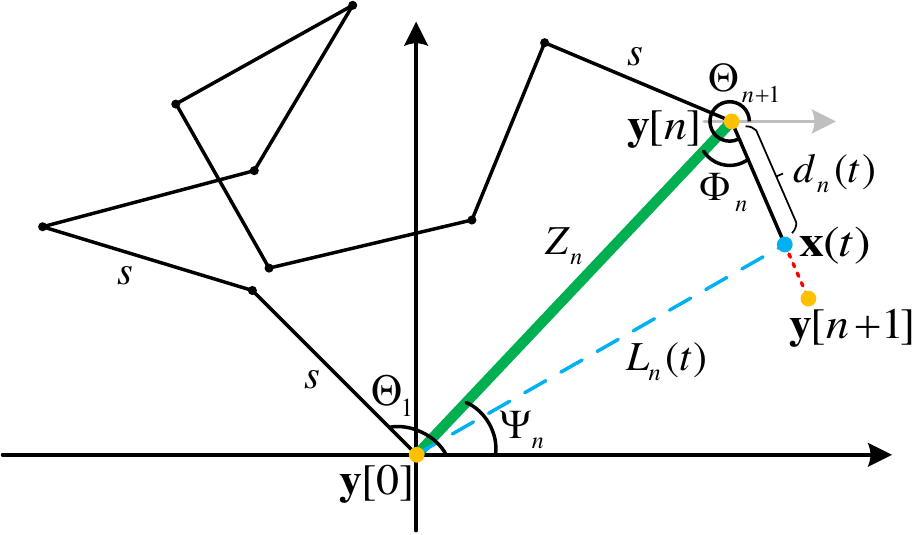}
\caption{A realization of the SRWP mobility model. The DBS is in its $(n+1)$-th flight and the red dotted line shows the rest of the $(n+1)$-th flight.}\vspace{-0.5cm}
\label{fig:SRWP}
\end{figure}

\begin{lemma} \label{lem:MainDensity}
In the UDM, for an interfering DBS located initially at $\nbx(0)$, let $L(t)$ be a random variable representing the distance from $\nbx(0)$ to the location of this interfering DBS at time $t$, and let the corresponding cumulative distribution function (cdf) and probability density function (pdf) be denoted by $F_L(l; t)$ and $f_L(l; t)$, respectively. Then $\Phi_{\rm D}'(t)$ will be an inhomogeneous PPP and its density is given as $\lambda(t; u_\nbx, u_0) =$
\begin{align} \label{MainLambda}
\scalebox{0.95}{$\lambda_0
  \begin{cases}
     1 & u_0 + vt \leq u_\nbx\\
     \beta(t, u_\nbx, u_0) & |u_0 - vt| \leq u_\nbx \leq u_0 + vt\\
     \beta(t, u_\nbx, u_0){\bf 1}\left(t>\frac{u_0}{v}\right) & 0 \leq u_\nbx \leq |u_0 - vt|
  \end{cases},$}
\end{align}
where ${\bf 1}(.)$ is the indicator function and
\begin{align} \label{MainBeta}
\beta&(t, u_\nbx, u_0) = 1 - F_L(u_0 - u_\nbx; t) \,- \nonumber\\
&\int_{|u_\nbx - u_0|}^{\min\{vt, u_\nbx + u_0\}} f_L(l; t)\frac{1}{\pi}\cos^{-1}\left(\frac{l^2 + u_\nbx^2 - u_0^2}{2lu_\nbx}\right)\,{\rm d}l.
\end{align}
\end{lemma}
\begin{IEEEproof}
See Appendix \ref{app:Lemma2}.
\end{IEEEproof}
According to Lemma \ref{lem:MainDensity}, the proper characterization of the network of interfering DBSs as they move based on the SRWP mobility model requires that we derive the distribution of $L(t)$, i.e., the displacement of each DBS at time $t$. Fig. \ref{fig:SRWP} shows a realization of the SRWP mobility model, where in each flight, DBSs hover for a constant time $w$ and then travel a constant distance $s$. Let $\nby[0]$ and $\nby[n]$ be the initial position of a DBS and its position after the $n$-th flight, respectively, and assume that $\Theta_n \sim U[0, 2\pi)$ is the angle between the direction of the $n$-th flight and the $x$-axis. Note that $\nby[n] = \nbx(t)$ for $n(w + \frac{s}{v}) \leq t < n(w + \frac{s}{v}) + w$, which corresponds to the hover time. We define $Z_n$ and $\Psi_n$ as the net displacement of a DBS between $\nby[0]$ and $\nby[n]$ and the angle between the $x$-axis and the line connecting $\nby[0]$ and $\nby[n]$, respectively. Furthermore, when a DBS is traveling in its $(n+1)$-th flight, we define $L_n(t) = \|\nbx(t) - \nbx(0)\|$ as the distance between $\nbx(0)$ and $\nbx(t)$. Note that when a DBS is in the $n$-th hover time, we have $L_n(t) = Z_n$. From Fig. \ref{fig:SRWP}, we observe that
\begin{align}
Z_n &= s\sqrt{\left(\sum_{i=1}^n \cos(\Theta_i)\right)^2 + \left(\sum_{i=1}^n \sin(\Theta_i)\right)^2},\label{SRWP_Z}\\
\Psi_n &= \tan^{-1}\left(\frac{\sum_{i=1}^n \sin(\Theta_i)}{\sum_{i=1}^n \cos(\Theta_i)}\right),\label{SRWP_Theta}\\
L_n(t) &= \sqrt{Z_n ^ 2 + d_n(t) ^ 2 - 2 Z_n d_n(t) \cos(\Phi_n)},\label{SRWP_L}
\end{align}
where $d_n(t) = vt - ns - (n + 1)vw$ is the distance traveled in the $(n + 1)$-th flight until time $t$, and $\Phi_n = \Theta_{n+1}-\Psi_n-\pi$ is the angle between the line connecting $\nby[0]$ and $\nby[n]$ and the direction of the $(n+1)$-th flight. Note that the pdf of $L(t)$ for the SRWP model is simply  $f_{Z_n}(l)$ when the DBS is hovering and $f_{L_n}(l; t)$ when the DBS is in motion. This can be compactly expressed as
\begin{align}\label{CDF_PDF_Lt}
&f_{L}(l; t) = \sum_{n=0}^\infty \! f_{Z_n}(l){\bf 1}\!\left(n(w \!+\! \frac{s}{v}) \leq t < n(w \!+\! \frac{s}{v}) \!+\! w\right) \,+\nonumber\\
&\hspace{0.4cm}\sum_{n=0}^\infty \! f_{L_n}(l; t){\bf 1}\!\left(n(w \!+\! \frac{s}{v}) \!+\! w \leq t < (n\!+\!1)(w \!+\! \frac{s}{v})\right).
\end{align}
We will characterize these two distributions in the rest of this section. The following lemma gives the distribution of $\Psi_n$, which is required in characterizing the distribution of $L_n(t)$.

\begin{lemma} \label{lem:RW_Psi_Uniform}
When interfering DBSs move based on the SRWP mobility model, the random variable $\Psi_n$ as defined in \eqref{SRWP_Theta}, is distributed uniformly in $[0, 2\pi)$. 
\end{lemma}
\begin{IEEEproof}
In order to derive the distribution of $\Psi_n$, we introduce $n - 1$ auxiliary random variables $\Psi_i = \Theta_i$, $1 \leq i \leq n - 1$, and find the joint pdf of $n$ random variables $\Psi_i$, $1 \leq i \leq n$.  We then integrate out these auxiliary random variables to find the pdf of $\Psi_n$. We start by solving the system of $n$ equations (one equation in \eqref{SRWP_Theta} and $n - 1$ equations introduced by the auxiliary random variables) to derive $\Theta_i$'s in terms of $\Psi_i$'s. The result can be written as two sets of solutions as follows:
\begin{align*}
{\rm Set~}1\!:
  \begin{cases}
	\Theta_i = \Psi_i, \hspace{0.5cm} i = 1, 2, \hdots, n - 1\\
	\Theta_n = \Psi_{n_1} = \tan^{-1} \left( \frac{-\Delta\cos(\Psi_n) + \Delta'\sin(\Psi_n)}{\Delta\sin(\Psi_n) + \Delta'\cos(\Psi_n)} \right),
  \end{cases}
\end{align*}
\begin{align*}
{\rm Set~}2\!:
  \begin{cases}
	\Theta_i = \Psi_i, \hspace{0.5cm} i = 1, 2, \hdots, n - 1\\
	\Theta_n = \Psi_{n_2} = \tan^{-1} \left( \frac{-\Delta\cos(\Psi_n) - \Delta'\sin(\Psi_n)}{\Delta\sin(\Psi_n) - \Delta'\cos(\Psi_n)} \right),
  \end{cases}
\end{align*}
where $\Delta =  \sum_{i=1}^{n-1} \sin(\Psi_i - \Psi_n)$ and $\Delta' = \sqrt{1 - \Delta^2}$. Writing the Jacobian matrix and computing its determinant, we get $|\ncalJ| = |\frac{\partial \Theta_n}{\partial \Psi_n}|$ for both solution sets. Hence, we have
\begin{align*}
|\ncalJ|\! =\! \frac{\sum_{i=1}^n\!\!\sum_{j=1}^n \!\!\cos(\Theta_i \!\!-\! \Theta_j)}{\sum_{i=1}^n \!\cos(\Theta_i \!\!-\! \Theta_n)} \!=\! 1 \!+\! \frac{\sum_{i=1}^n\!\!\sum_{j=1}^{n-1}\!\!\cos(\Theta_i \!\!-\! \Theta_j)}{\sum_{i=1}^n \!\cos(\Theta_i \!\!-\! \Theta_n)}.
\end{align*}
By some algebraic manipulations, we find that $|\ncalJ|\bigr\rvert_{\Theta_n = \Psi_{n_1}} + |\ncalJ|\bigr\rvert_{\Theta_n = \Psi_{n_2}} = 2$. Now, since $\Theta_i \sim U[0, 2\pi)$, the joint distribution of $\Psi_i$'s can be written as
\begin{align*}
f_{\boldsymbol {\Psi}}(\boldsymbol {\psi}) \!=\! f_{\boldsymbol {\Theta}}(\boldsymbol {\theta}) |\ncalJ|\bigr\rvert_{\Theta_n = \Psi_{n_1}} \!\!\!\!+\! f_{\boldsymbol {\Theta}}(\boldsymbol {\theta}) |\ncalJ|\bigr\rvert_{\Theta_n = \Psi_{n_2}} \!\!\!=\! 2\left(\frac{1}{2\pi}\right)^n,
\end{align*}
where the boldface letters represent vector random variables. Integrating out $\Psi_i$, $1 \leq i \leq n - 1$, the distribution of $\Psi_n$ is derived as $f_{\Psi_n}(\psi_n) = \frac{1}{\pi}$ for $\psi_n \in [-\frac{\pi}{2}, \frac{\pi}{2})$, due to the range of $\tan^{-1}$ function. Finally, since the true range of $\Psi_n$ is $[-\pi, \pi)$, we conclude that $f_{\Psi_n}(\psi_n) = \frac{1}{2\pi}$ for $\psi_n \in [-\pi, \pi)$, and the proof is complete. 
\end{IEEEproof}
\begin{remark} \label{Remark_Arcsine}
Since $\Theta_{n+1}$ and $\Psi_n$ are independent and identically distributed (i.i.d.) and uniform in $[0, 2\pi)$, we conclude that $\Phi_n$ will have a symmetric triangular distribution. However, since the range of values of $\Phi_n$ is between $0$ and $2\pi$, we have $\Phi_n \sim U[0, 2\pi)$. 
\end{remark}
We can find the distribution of $Z_n$ using the same method as in the proof of Lemma \ref{lem:RW_Psi_Uniform}. However, it turns out that this distribution consists of an $n$-fold integral with no closed-form solution. On the other hand, we can characterize an asymptotic distribution for $Z_n$ as $n\to \infty$. This result is given in the next lemma.
\begin{lemma} \label{lem:RW_Z_Rayleigh_Asymptotic}
When interfering DBSs move based on the SRWP mobility model, the random variable $Z_n$ as defined in \eqref{SRWP_Z}, will have a Rayleigh distribution with parameter $s\sqrt{\frac{n}{2}}$, i.e., $f_{Z_n}(z) = \frac{2z}{ns^2}{\rm e}^{-\frac{z^2}{ns^2}}$ as $n \to \infty$.
\end{lemma}
\begin{IEEEproof}
Define $X \!=\! \sum_{i=1}^n \cos(\Theta_i)$ and $Y \!=\! \sum_{i=1}^n \sin(\Theta_i)$. Since $\Theta_i$'s are i.i.d. with uniform distribution in $[0, 2\pi)$, the central limit theorem (CLT) asserts that as $n \to \infty$, $X$ and $Y$ will have Gaussian distributions. For the moments of $X$ we have $\nbbE[X] = \sum_{i=1}^n \nbbE[\cos(\Theta_i)] = 0$ and $\nbbE[X ^ 2] = \nbbE\!\left[\sum_{i=1}^n\!\sum_{j=1}^n \!\!\cos(\Theta_i)\!\cos(\Theta_j)\right] \!= \!\nbbE\!\left[\sum_{i=1}^n  \!\!\cos ^ 2(\Theta_i)\right] \!=\! \frac{n}{2}$.
Note that the same is also true for $Y$. Hence, $X\! \sim \!\ncalN(0,  \frac{n}{2})$ and $Y \!\sim\! \ncalN(0,  \frac{n}{2})$. Now since $\nbbE[XY] = 0$, we conclude that $X$ and $Y$ are uncorrelated, and thus, independent. Therefore $Z_n \!=\! s\sqrt{X^2 \!+\! Y^2}$ will have a Rayleigh distribution with parameter $\sigma = s\sqrt{\frac{n}{2}}$ as $n \to \infty$ and the proof is complete.
\end{IEEEproof}
Using the asymptotic distribution of $Z_n$, we can approximate the distribution of $Z_n$ with a Rayleigh distribution. However, since in the SRWP mobility model we already know that $Z_n \leq ns$, a more appropriate choice for the approximate distribution of $Z_n$ is the truncated Rayleigh distribution, which is defined over the interval $[0, A]$ with scale parameter $\sigma$ by the following pdf:
\begin{align}
f(x;A,\sigma) = \frac{\frac{x}{\sigma^2}{\rm e}^{-\frac{x^2}{2\sigma^2}}}{1 - {\rm e}^{-\frac{A^2}{2\sigma^2}}}{\bf 1}\left( 0\leq x \leq A \right).
\end{align}
Note that the distribution of $Z_n$ for $n = 1$ is trivial and for $n=2$ one can easily show that it follows an arcsine distribution. Hence, we have the following approximation:
\begin{align}\label{PDF_Zn}
f_{Z_n}(z) \approx
  \begin{cases}
     \delta(z - s) & n=1\\
     \frac{2}{\pi\sqrt{(2s)^2 - z^2}}{\bf 1}\left( 0\leq z \leq 2s \right) & n=2\\
     \frac{2z{\rm e}^{-\frac{z^2}{n s^2}}}{n s^2(1 - {\rm e}^{-n})}{\bf 1}\left( 0\leq z \leq n s \right) & n \geq 3
  \end{cases},
\end{align}
where $n=k$ implies that $k(\frac{s}{v} + w) \leq t \leq k(\frac{s}{v} + w) + w$.

Getting back to \eqref{SRWP_L}, we can now compute the distribution of $L_n(t)$ for a given time $t$ as follows.
\begin{align}\label{CDF_Ln}
F_{L_n}(l;t) &= \nbbP[Z_n ^ 2 + d_n(t) ^ 2 - 2 Z_n d_n(t) \cos(\Phi_n) \leq l^2] \nonumber\\
&\overset{(a)}{=} \!\!\int_0^{ns}\!\! \nbbP\left[ \cos\left(\Phi_n\right) \geq \frac{z ^ 2 + d_n(t) ^ 2 - l ^ 2}{2 z d_n(t)} \right]\!f_{Z_n}(z)\,{\rm d}z\nonumber\\
&\overset{(b)}{=} F_{Z_n}(l-d_n(t)) \,+ \nonumber\\
&\hspace{-1cm}\int_{|l-d_n(t)|}^{\min\{l + d_n(t), ns\}}\!\frac{1}{\pi} \cos^{-1}\left(\frac{z ^ 2 + d_n(t) ^ 2 - l ^ 2}{2 z d_n(t)}\right)\!f_{Z_n}(z)\,{\rm d}z,
\end{align}
where in $(a)$ we conditioned the probability on knowing $Z_n$ and in $(b)$ we used Remark \ref{Remark_Arcsine} with some mathematical manipulations to simplify the result. Taking the derivative of \eqref{CDF_Ln} with respect to $l$, we can write the pdf of $L_n(t)$ as
\begin{align}\label{PDF_Ln}
f_{L_n}(l;t)\! = \!\!\!\int_{|l-d_n(t)|}^{\min\{l + d_n(t), ns\}}\hspace{-1.1cm}\frac{2l \,f_{Z_n}(z) ~{\rm d}z}{\pi\sqrt{(l^2 \!-\! (z\!-\!d_n(t))^2)((z\!+\!d_n(t))^2 \!-\! l^2)}}.
\end{align}
Hence, the approximate distribution of $L(t)$ can be obtained by inserting \eqref{PDF_Zn} and \eqref{PDF_Ln} into \eqref{CDF_PDF_Lt}. Finally, the density of the interference field in the UDM is derived by applying \eqref{CDF_PDF_Lt} to Lemma \ref{lem:MainDensity}.

\vspace{-0.3cm}
\section{Average Rate} \label{sec:Rate}
Having derived the density of the network of interfering DBSs for both the UIM and the UDM, we can now compute the average rate achieved by the typical UE at time $t$. The result is provided in the next theorem.
\begin{theorem} \label{theo:ASE}
In the UDM, the average rate achieved by the typical UE at time $t$ can be written as
\begin{align}
R(t) &= \int_0^\infty \int_0^\infty \frac{2\pi\lambda_0 u_0 {\rm e}^{-\pi \lambda_0 u_0^2}}{1+\gamma} \,\times \nonumber\\
&\hspace{-0.5cm}\exp\Big(- 2\pi{\displaystyle \int_{0}^\infty}\frac{u_\nbx \lambda(t; u_\nbx, u_0)}{1 + \frac{1}{\gamma}\left(\frac{u_\nbx^2+h^2}{u_0^2(t)+h^2}\right)^{\alpha/2}}\,{\rm d}u_\nbx\Big) \,{\rm d}u_0, \label{MainRates}
\end{align}
where $\lambda(t; u_\nbx, u_0)$ is given in Lemma \ref{lem:MainDensity} and $u_0(t) = [u_0 - vt]^+$.
\end{theorem}
\begin{IEEEproof}
We start by writing the complementary cumulative distribution function (ccdf) of ${\rm SIR(t)}$ conditioned on the location of the serving DBS as
\begin{align*}
\nbbP\!\left[{\rm SIR}(t)\! \geq\! \gamma \bigr\rvert\nbx_0(t) \right]\!\! &\overset{(a)}{=}\! \mathbb{E}\!\left[\nbbP\!\left[ h_0(t) \!\geq\! \frac{\gamma r_0^\alpha(t)I(t)}{P}\middle|\nbx_0(t),\! I(t) \right]\right]\nonumber\\
&\overset{(b)}{=} \ncalL_{I(t)}\left(s\bigr\rvert\nbx_0(t)\right)\biggr|_{s=\frac{\gamma r_0^\alpha(t)}{P}},
\end{align*}
where in $(a)$ the expectation is taken over $I(t)$ and in $(b)$ the Rayleigh fading assumption is used and $\ncalL_{I(t)}(s\bigr\rvert\nbx_0(t)) = \nbbE\left[{\rm e}^{-sI(t)}\bigr\rvert\nbx_0(t)\right]$ represents the conditional Laplace transform of interference at time $t$, which can be computed as
\begin{align*}
\ncalL_{I(t)}(s\bigr\rvert\nbx_0(t)) &= \nbbE\left[\exp\left[-s\!\!\!\!\!\!\!\!\sum_{\nbx(t) \in \Phi_{\rm D}'(t)}\!\!\!\!\!\!\!\!\!P h_\nbx(t) r_\nbx(t)^{-\alpha}\right] \middle|u_0(t) \right] \nonumber\\
&\hspace{-0.8cm}\overset{(a)}{=} \nbbE\left[ \prod_{\nbx(t) \in \Phi_{\rm D}'(t)} \frac{1}{1 + sP(u_\nbx^2(t)+h^2)^{-\alpha/2}} \middle|u_0(t) \right] \nonumber\\ 
&\hspace{-0.8cm}\overset{(b)}{=} \exp\left[-2\pi\int_{0}^\infty\frac{u_\nbx(t) \lambda(t; u_\nbx, u_0)}{1 + \frac{1}{sP}(u_\nbx^2(t)+h^2)^{\alpha/2}}\,{\rm d}u_\nbx(t)\right],
\end{align*}
where (a) results from the moment generating function (MGF) of the exponential distribution and (b) follows from the probability generating functional (PGFL) of a PPP. Now, we can write the average rate at time $t$ as
\begin{align*}
R(t) &= \nbbE[\log\left(1 + {\rm SIR}(t)\right)] = \int_0^\infty \log(1+\gamma)f_\Gamma(\gamma; t)\,{\rm d}\gamma\\
&= \int_0^\infty \!\!\!\! \int_0^\infty \frac{2\pi\lambda_0 u_0 {\rm e}^{-\pi \lambda_0 u_0^2}}{1+\gamma}\nbbP\!\left[{\rm SIR}(t)\! \geq\! \gamma \bigr\rvert\nbx_0(t) \right]{\rm d}u_0\,{\rm d}\gamma,
\end{align*}
where $f_\Gamma(\gamma; t)$ is the pdf of ${\rm SIR}(t)$ and in the last equation we used integration by parts and deconditioned on $u_0(t)$. This completes the proof.
%
%
\end{IEEEproof}
For the UIM, since the density of the network of interfering DBSs is given in \eqref{LambdaServiceModel1}, the received rate at the typical UE will be \eqref{MainRates} evaluated at $t=0$, i.e.,
\begin{align}
R &= \int_0^\infty \int_0^\infty \frac{2\pi\lambda_0 u_0 {\rm e}^{-\pi \lambda_0 u_0^2}}{1+\gamma} \times \nonumber\\
&\hspace{-0.1cm}\exp\Big(- 2\pi\lambda_0{\displaystyle \int_{u_0}^\infty}\frac{u_\nbx}{1 + \frac{1}{\gamma}\left(\frac{u_\nbx^2+h^2}{u_0^2+h^2}\right)^{\alpha/2}}\,{\rm d}u_\nbx\Big) \,{\rm d}u_0 \,{\rm d}\gamma. \label{SubMainCoverage}
\end{align}

\begin{figure}[t!]
    \centering
    \includegraphics[width=0.8\columnwidth]{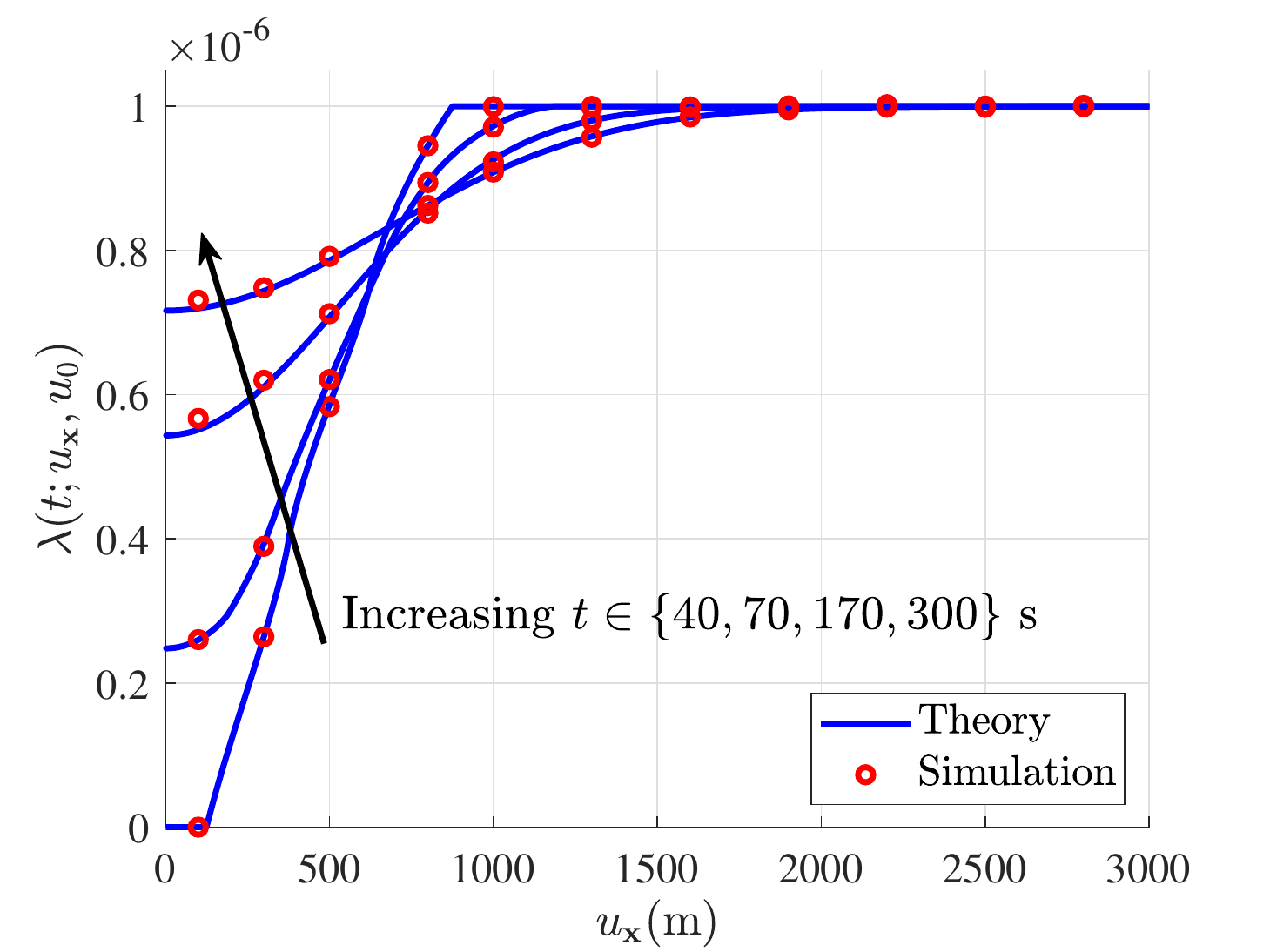}
    \caption{Density of the network of interfering DBSs for the UDM where $u_0 = 500~{\rm m}$. The accuracy of our approximations is evident in this figure.}\vspace{-0.3cm}
    \label{Fig:DensityPlot}
\end{figure}

\vspace{-0.1cm}
\section{Numerical Results} \label{sec:Simulation}
In this section, we perform simulations to verify the accuracy of the key approximations and glean useful insights about the system-level performance. We assume that the DBSs are distributed initially as a PPP with $\lambda_0 = 10 ^ {-6}$ on the DBS plane. The DBSs are assumed to move at a constant velocity of $v = 45~{\rm km/h}$ based on the SRWP mobility model, in which $w = 5$ seconds and $s = 250$ meters. We assume $\alpha = 3$ and $h \in \{100, \, 150, \, 200\}$ meters.

Fig. \ref{Fig:DensityPlot} presents the density of the network of interfering DBSs for the UDM, where $u_0 = 500$ meters and $t \in \{40, 70, 170, 300\}$ seconds. From this figure, it is clear that our truncated Rayleigh approximation is quite accurate. Moreover, as $t\to\infty$, the interference field will become homogeneous.

\begin{figure}[t!]
    \centering
    \includegraphics[width=0.8\columnwidth]{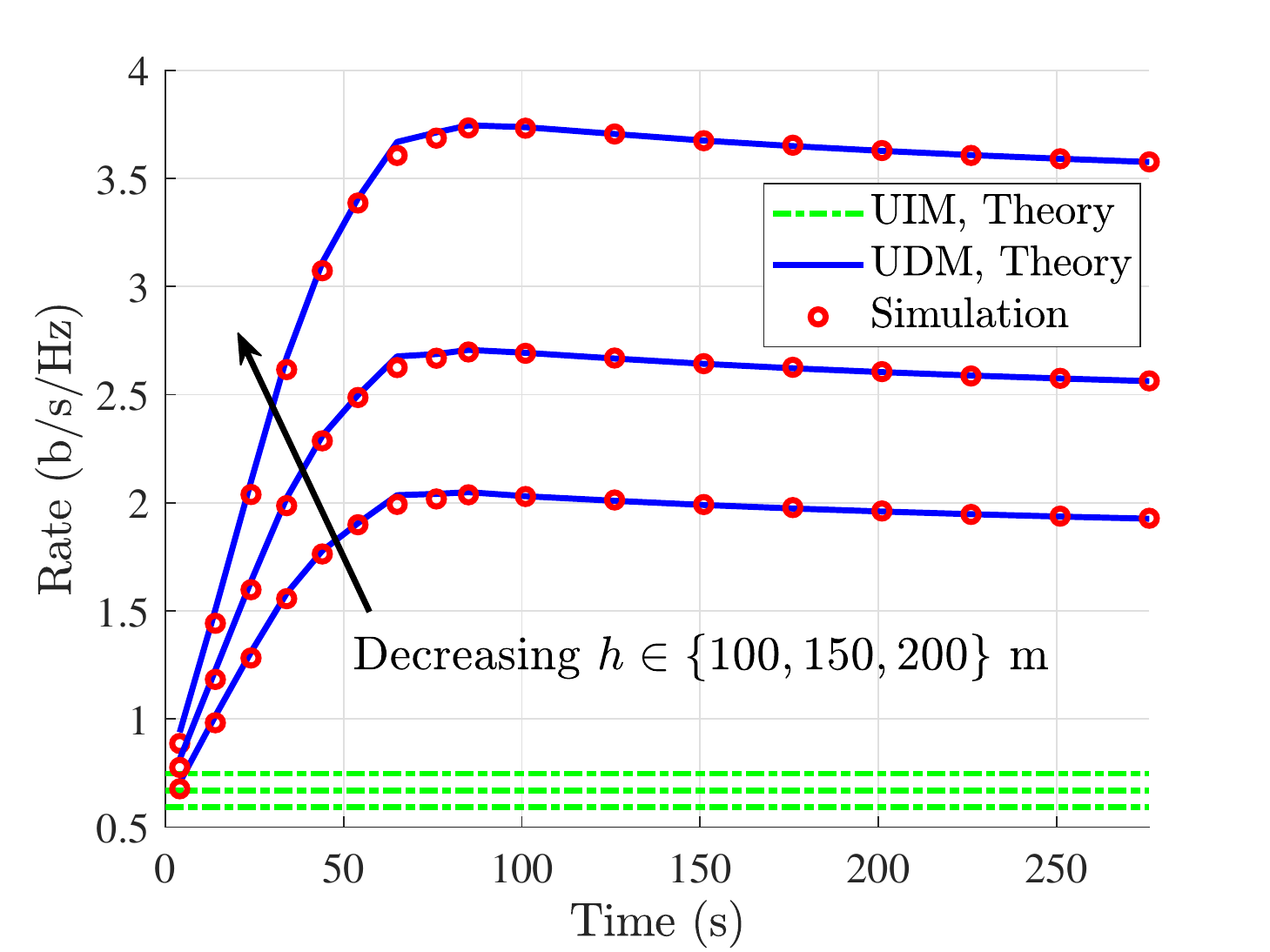}
    \caption{Average rate achieved by the typical UE at time $t$ for both the UIM and the UDM. DBSs are moving at $h \in \{100, 150, 200\}$ meters and $\alpha = 3$.}\vspace{-0.3cm}
    \label{Fig:RatePlot_VaryHeight}
\end{figure}

In Fig. \ref{Fig:RatePlot_VaryHeight}, we show the average rate as a function of time for both the UIM and the UDM at various heights. As can also be inferred from Theorem \ref{theo:ASE}, the received rate at the typical UE will decrease as the height increases. The plots for the UIM are given in this figure in order to showcase the advantage of the UDM (in which the trajectories are UE dependent) over the UIM.

\vspace{-0.15cm}
\section{Conclusion} \label{sec:Conclusion}
In this paper, we have performed a comprehensive analysis on a network of mobile DBSs that are serving UEs on the ground. Assuming that initial locations of the drones follow a homogeneous PPP and the nearest neighbor association policy is used for determining the serving DBS, we proposed two service models for the serving DBS, i.e., (i) the serving DBS moves based on the SRWP independently of the typical UE (UIM), and (ii) the serving DBS moves towards the typical UE and keeps hovering above its location until its transmission to the typical UE is completed (UDM). All the other DBSs are treated as interfering DBSs, whose mobility is described by the SRWP model. We then characterized several fundamental properties of the SRWP mobility model using which we analyzed the interference field as seen by the typical UE for both the UIM and the UDM. Finally, we computed the average rate at the typical UE as a function of time for both service models. To the best of our knowledge, this is the first work that performs a concrete analysis on the performance of a mobile drone network, where the mobility of the drones is described by an SRWP mobility model on an infinite plane. It should also be noted that this non-linear mobility model is a significant generalization of the ``straight-line" mobility model used by the standardization bodies, such as 3GPP.


\vspace{-0.1cm}
\appendix
\subsection{Proof of Lemma \ref{lem:MainDensity}} \label{app:Lemma2}
For the network of interfering DBSs, since we have started with an inhomogeneous PPP of the initial density given in \eqref{LambdaServiceModel1} and the displacements are independent of each other, displacement theorem asserts that the resulting network will also be an inhomogeneous PPP \cite{B_Haenggi_Stochastic_2012}. According to Lemma \ref{lem:NoExclusion}, if there was no exclusion zone $\ncalX$, the density of the DBS network (including the serving DBS) would be $\lambda_0$ as they move independently based on the SRWP mobility model. On the other hand, with $\ncalX$, the resulting density of the network can be partitioned into two parts: (i) density due to the points initially inside $\ncalX$ (denoted as $\lambda_1(t; u_\nbx, u_0)$), and (ii) density due to the points initially outside $\ncalX$ (denoted as $\lambda(t; u_\nbx, u_0)$). Note that the latter determines the density of the network of interfering DBSs. Therefore, we have $\lambda(t; u_\nbx, u_0) = \lambda_0 - \lambda_1(t; u_\nbx, u_0)$.



Let $N(t)$ be the average number of points that are inside $\ncalX$ at $t=0$ and land on an infinitesimal annulus with an inner and outer radii of $u_\nbx$ and $u_\nbx+ {\rm d}u_\nbx$, respectively, after a displacement of $L(t)$. Note that due to the symmetry of the network, the density is rotation invariant, and thus, it is sufficient to consider an annulus for our analysis. We have
%
\begin{align} \label{DefLambda}
\lambda_1(t; u_\nbx, u_0) &= \lim_{{\rm d}u_\nbx \to 0}\frac{N(t)}{2\pi u_\nbx {\rm d}u_\nbx}.
\end{align}


\begin{figure}
\centering
\includegraphics[width=0.5\columnwidth]{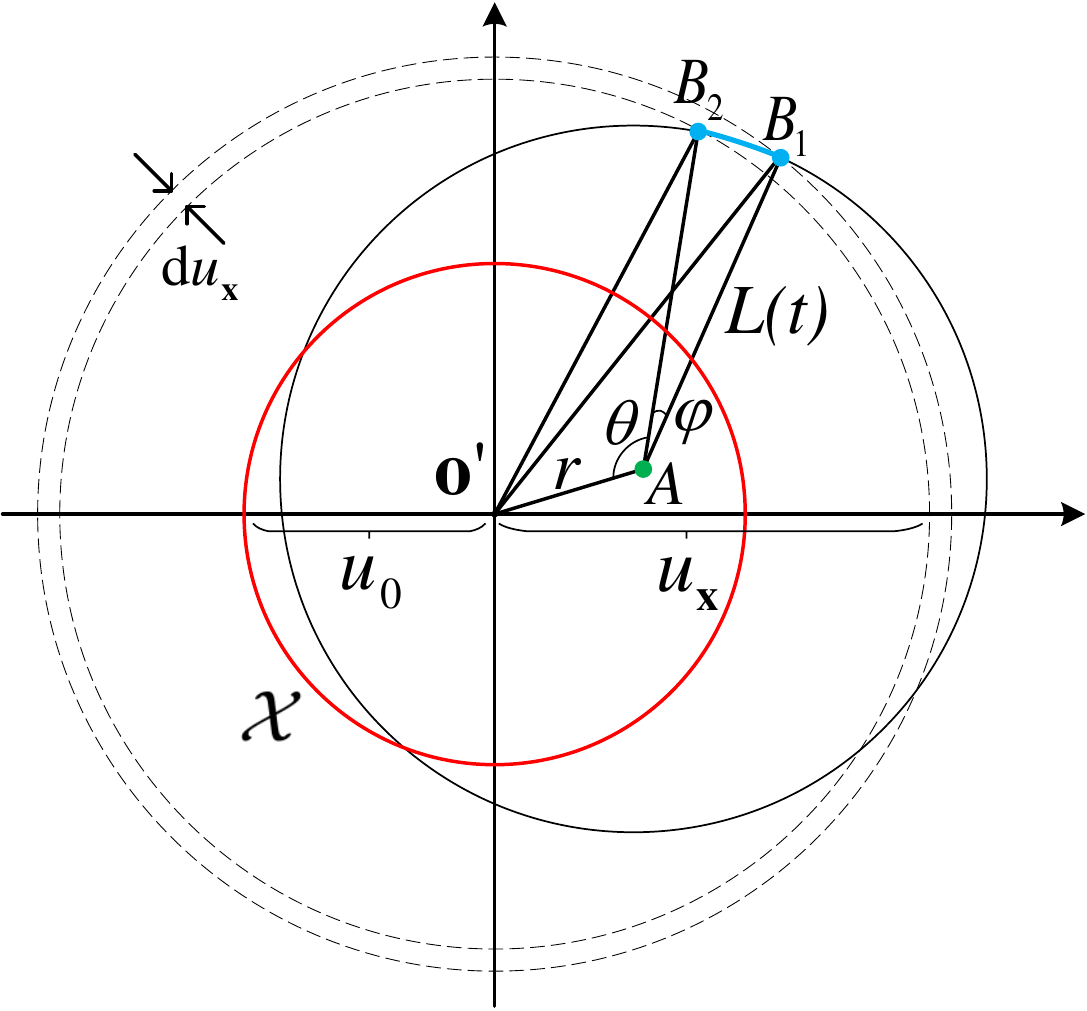}
\caption{An illustration for the proof of Lemma \ref{lem:MainDensity}. The red circle indicates $\ncalX$ and dotted circles denote the annulus of interest.}
\vspace{-0.5cm}
\label{fig:ProofTheorem1_1}
\end{figure}
As depicted in Fig. \ref{fig:ProofTheorem1_1}, we represent a randomly selected point inside $\ncalX$ by $A$, which has a distance of $r$ from ${\bf o'}$. Denote the two intersection points of $b(A, L(t))$ with the annulus of interest by $B_1$ and $B_2$. Writing the cosine law in triangles ${{\bf o'}AB_2}$ and ${{\bf o'}AB_1}$, we have
\begin{align*}
u_\nbx^2 &= L(t)^2 + r^2 - 2rL(t)\cos(\theta)\\
(u_\nbx + {\rm d}u_\nbx)^2 &= L(t)^2 + r^2 - 2rL(t)\cos(\theta + \varphi),
\end{align*}
where $\varphi = \angle B_1AB_2$ and $\theta = \angle{\bf o'}AB_2$. Eliminating $\theta$ from these equations yields
\begin{align*}
\cos(\varphi) &= \frac{1}{b^2}\Bigg[a^2 - a\left({\rm d}u_\nbx + \frac{({\rm d}u_\nbx)^2}{2u_\nbx}\right) + \nonumber\\
&c\sqrt{c^2 + 2a\left({\rm d}u_\nbx + \frac{({\rm d}u_\nbx)^2}{2u_\nbx}\right) - \left({\rm d}u_\nbx + \frac{({\rm d}u_\nbx)^2}{2u_\nbx}\right)^2}\Bigg],
\end{align*}
where $a = \frac{L(t)^2 + r^2 - u_\nbx^2}{2u_\nbx}, b = \frac{2rL(t)}{2u_\nbx}, c = \sqrt{b^2 - a^2}$.
Note that we cannot ignore the effect of terms having $({\rm d}u_\nbx)^2$ under the square root function, because they will ultimately contribute to the final limiting behavior of the function.

Since the direction of movement of DBSs is uniformly distributed from $0$ to $2\pi$, the probability that a DBS at point $A$ lands on the annulus of interest after the displacement of $L(t)$ is $\varphi / \pi$. Now, $N(t)$ can be written by taking all such points in $\ncalX$ into account and then averaging over the random variable $L(t)$. 
Hence, $N(t) = \nbbE\left[\int_0^{u_0} \frac{\varphi}{\pi} 2\pi r \lambda_0\,{\rm d}r\right]$, where the expectation is taken over $L(t)$. This gives $\lambda_1(t; u_\nbx, u_0)$ as
\begin{align}\label{EqLim1}
\lambda_1(t; u_\nbx, u_0) &= \frac{\lambda_0}{\pi}\nbbE\left[\int_0^{u_0} \frac{r}{u_\nbx} \lim_{{\rm d}u_\nbx \to 0}\frac{\varphi}{{\rm d}u_\nbx}\,{\rm d}r\right],
\end{align}
where we changed the order of limit with expectation and integration. Denoting ${\rm d}u_\nbx$ as $x$ for simplicity, we can compute the limit as follows.
\begin{align}\label{EqLim2}
\lim_{x \to 0}\frac{\varphi}{ x} &= \lim_{x \to 0}\frac{1}{x}\cos^{-1}\Bigg(1 - \frac{1}{a^2 + c^2}\Bigg[c^2 + ax + \frac{a}{2 u_\nbx}x^2 \,- \nonumber\\
&\hspace{1.5cm}c\sqrt{c^2 + 2ax + \frac{a}{ u_\nbx }x^2 - \left( x + \frac{x^2}{2 u_\nbx } \right)^2}\Bigg]\Bigg)\nonumber\\
&\hspace{-0.5cm}\overset{(*)}= \!\!\sqrt{\frac{2}{a^2 + c^2}}\!\lim_{x \to 0}\sqrt{\frac{(a^2 + c^2)\left(x^2 + \frac{1}{ u_\nbx }x^3 + \frac{1}{4 u_\nbx ^2}x^4\right)}{2c^2x^2}}\nonumber\\
&\hspace{-0.5cm}= \frac{1}{c} = \frac{2 u_\nbx}{\sqrt{( u_\nbx ^2 - (L(t)-r)^2)((L(t)+r)^2 -  u_\nbx ^2)}},
\end{align}
where in $(*)$ we used the Taylor series expansion $\cos^{-1}(1-x) = \sqrt{2x} + \Theta(x^{3/2})$ as $x\to 0$, where $p(t) = \Theta(q(t))$ implies that $p(t)$ is asymptotically bounded by $q(t)$ from both above and below. Note that since the triple $(u_\nbx, r, L(t))$ form a triangle, the result in \eqref{EqLim2} is real and positive, as expected. Plugging \eqref{EqLim2} back into \eqref{EqLim1}, we have $\lambda_1(t; u_\nbx, u_0) =$
\begin{align} \label{Lambda1First}
\frac{\lambda_0}{\pi}{\int_0^{\infty}}\int_{\ncalR_1}\frac{2r f_L(l;t)}{\sqrt{(u_\nbx^2 - (l-r)^2)((l+r)^2 - u_\nbx^2)}}\,{\rm d}r\,{\rm d}l,
\end{align}
where $\ncalR_1 = \left\{|l-u_\nbx| \leq r \leq l+u_\nbx\right\} \bigcap \left\{0 \leq r \leq u_0\right\}$. To simplify \eqref{Lambda1First}, recall from the definition of $L(t)$ that the inequality $L(t) \leq vt$ always holds, since the net displacement of a DBS at time $t$ cannot exceed the total distance traveled by the DBS, i.e., $vt$. Hence, \eqref{Lambda1First} simplifies to $\lambda_1(t; u_\nbx, u_0) =$
\begin{align*} 
&\frac{\lambda_0}{\pi}{\int_0^{vt}}\!\!\!\int_{\ncalR_1}\!\! f_L(l;t)\frac{2r}{\sqrt{(u_\nbx^2 \!-\! (l-r)^2)((l+r)^2 \!-\! u_\nbx^2)}}\,{\rm d}r\,{\rm d}l~+\nonumber\\
&\frac{\lambda_0}{\pi}(1\!-\!F_L(vt;t))\!\!\int_{\ncalR_2}\!\!\frac{2r}{\sqrt{(u_\nbx^2 \!-\! (vt-r)^2)((vt+r)^2\! -\! u_\nbx^2)}}\,{\rm d}r,
\end{align*}
where $\ncalR_2 = \left\{|vt-u_\nbx| \leq r \leq vt+u_\nbx\right\} \bigcap \left\{0 \leq r \leq u_0\right\}$. Simplifying the last step requires careful integrations and the details are omitted here for brevity. Finally, the density of the network of interfering DBSs is summarized as equations \eqref{MainLambda} and \eqref{MainBeta} in the lemma statement. This completes the proof.
\hfill 
\IEEEQED
\bibliographystyle{IEEEtran}
\bibliography{C2_RWP_Drone}

\end{document}